\documentclass[10pt, conference]{IEEEtran}

\usepackage{amssymb}
\usepackage{amsmath}

\usepackage[colorlinks=true, linkcolor=blue, citecolor=blue, urlcolor=blue]{hyperref}

\usepackage{graphicx}
\usepackage{booktabs}
\usepackage{multirow}
\usepackage{tabularx}
\usepackage{enumitem}
\usepackage{threeparttable}
\usepackage{balance}
\usepackage{url}

\usepackage{fancyhdr} % for arxiv
\fancypagestyle{firststyle}
{
\fancyhf{}
\fancyfoot[C]{\scriptsize{Proceedings of the 22nd Annual International Conference on Privacy, Security, and Trust (PST 2025), Fredericton, IEEE, pp.~1--10. \\ This version is the authors' copy. The publisher's definite version is available via \url{https://doi.org/10.1109/PST65910.2025.11268853}.}}}

\begin{document}

\title{A Static Analysis of \\ Popular C Packages in Linux}

\author{
\IEEEauthorblockN{Jukka Ruohonen}
\IEEEauthorblockA{University of Southern Denmark \\
Email: juk@mmmi.sdu.dk}
\and
\IEEEauthorblockN{Mubashrah Saddiqa}
\IEEEauthorblockA{University of Southern Denmark \\
Email: msad@mmmi.sdu.dk}
\and
\IEEEauthorblockN{Krzysztof Sierszecki}
\IEEEauthorblockA{University of Southern Denmark \\
Email: krzys@mmmi.sdu.dk}
}

\maketitle

\begin{abstract}
Static analysis is a classical technique for improving software security and software quality in general. Fairly recently, a new static analyzer was implemented in the GNU Compiler Collection (GCC). The present paper uses the GCC’s analyzer to empirically examine popular Linux packages. The dataset used is based on those packages in the Gentoo Linux distribution that are either written in C or contain C code. In total, 3,538 such packages are covered. According to the results, uninitialized variables and NULL pointer dereference issues are the most common problems according to the analyzer. Classical memory management issues are relatively rare. The warnings also follow a long-tailed probability distribution across the packages; a few packages are highly warning-prone, whereas no warnings are present for as much as 89\% of the packages. Furthermore, the warnings do not vary across different application domains. With these results, the paper contributes to the domain of large-scale empirical research on software quality and security. In addition, a discussion is presented about practical implications of the results.
\end{abstract}


\section{Introduction}

\thispagestyle{firststyle} % for arxiv

With Linux serving as the foundation for many critical systems, from servers to
embedded devices, ensuring its security and code quality is of high
importance. Among the many techniques employed to enhance and improve software
robustness, static analysis is a proactive approach for detecting defects,
including security vulnerabilities, early on in the software development
processes. With this backdrop in mind, the paper presents a large-scale
empirical analysis of warnings outputted by the GCC's relatively new static
analyzer for popular open source software packages entirely or partially written
in the classical C programming language. The paper is the first to use the GCC's
analyzer for research purposes. In addition, the paper contributes to the domain
of large-scale empirical studies on software security and software quality,
offering also some ideas on how improvements could be made in practice,
including with respect to the static analyzer and the open source software
domain in general.

In software testing it is common to make a distinction between dynamic and
static analysis. The former includes various techniques to detect problems
during run-time of a program; the examples include unit testing, dynamic
symbolic execution, performance analysis, and so-called fuzzing, which has
become popular also in the open source context~\text{\cite{Ding21,
    Ruohonen19RSDA}}. In contrast, static analysis tests a program without
executing it. The history of static analysis traces all the way back to the
1970s during which first static analyzers were implemented based on the 1950s
and 1960s ideas about compiler design~\cite{Thomson21}. Depending on whether
source code or binary code is tested, static analyzers range from simple
lint-like checkers to more complex solutions based on abstract syntax trees,
control flow graphs, model checking, and other techniques. A further taxonomy is
between user-oriented and developer-oriented software
testing~\cite{Hjerppe19IFIP}. The present work belongs to the developer-oriented
domain; the paper's topic is about software verification, although the actual
verification is still only implicit, drawing on empirical observations.

Static analysis has long been important in software development for improving
code quality and software security. The benefits include: swiftness compared to
manual code reviews and dynamic analysis, complete and consistent coverage, and
ease for non-experts to review code~\cite{West08}, among other things. According
to surveys, static analysis is indeed often used in conjunction with code
reviews~\cite{Ryan23a}. Against this background, it could be argued that static
analysis is particularly relevant for open source software
projects~\cite{Alexopoulos20}, given the long-lasting debate over the
effectiveness of code reviews in the open source context~\cite{Payne02}. Though,
it must be stressed that static analysis cannot expel other testing techniques
and human expertise. Rather, static analysis complements these and other means
to improve software quality, including software security.

A static analyzer is presumably easier to implement for strongly typed
programming languages than for weakly typed ones. Although C is only partially
strongly typed, given a number of implicit conversions between types, pointer
arithmetic, and aliasing, static analysis is particularly relevant for C already
because of the language's notoriety in increasing the likelihood of difficult
bugs, including various software vulnerabilities, such as buffer and integer
overflows. In fact, some open source coding style guidelines and related guides
for C recommend turning all compiler warnings into errors, such that a program
does not successfully compile before a given warning is fixed~\cite{OpenSSF24}.
Even though such practices may not be enforced in all open source
projects~\cite{Beller16, Vassallo20}, the recent integration of static analyzers
into continuous integration pipelines of many open source projects has likely
eased the enforcement and associated coding
practices~\cite{Zampetti17}. Recently, a lot of effort has also been devoted for
improving the static analysis capabilities of C compilers, thus helping
developers to avoid common mistakes. This empirical paper relies on these new
capabilities for examining problems found through static analysis of popular
Linux packages written in C.

Before continuing any further, a remark about terminology is in order: there is
no universally accepted term for describing the ``problems'' found via static
analysis. These have been referred to as issues~\cite{Ruohonen21PST},
smells~\cite{Fard13, Molnar24}, \text{warnings~\cite{Guo23, Molnar24}}, or even
bugs~\cite{Ayewah08, Molnar24}, although smells, warnings, and bugs all require
further deducing. In other words, a given ``problem'' found through static
analysis may not amount to an actual bug or a smell. It may even be that a
``problem'' is not really a problem; false positives are common. To make things
explicit and to avoid further confusion, in what follows, a term ``compile-time
warning'' (CTW) is~used. Implicitly, these CTWs are assumed to map to software
weaknesses, as described in the Common Weakness Enumeration (CWE) maintained by
the MITRE corporation~\cite{MITRE24a}. These weaknesses are about security; a
CWE may manifest itself as a concrete software vulnerability, although again
further verification is required to make the connection. The paper neither
considers such a verification nor deduces about the CTW$\mapsto$CWE mappings in
detail. As described in Section~\ref{sec: methods and data}, the mappings are
taken for granted based on the static analyzer used. Thus, a~grain of salt
should be taken about the security-orientedness of the results presented, but
still only a grain; by assumption, the analyzer is fairly robust in this regard.

On these notes, the three research questions (RQs) examined can be stated as follows:
\begin{itemize}
\itemsep 5pt
\item{RQ.1: How common are CWE-mapped CTWs in popular Linux packages fully or
  partially written in the C programming language?}
\item{RQ.2: Which CWEs are particularly common according to the GCC's compile-time warnings?}
\item{RQ.3: Do the CWE-mapped CTWs vary across typical application domains, such
  as system libraries, networking tools, or graphical user interfaces?}
\end{itemize}

These three research questions carry novelty. According to a reasonable
literature search, as well as related arguments raised in the
literature~\cite{Guo23}, the questions have not been previously examined, at
least to the large-scale extent presented in the current work. Nor has the GCC's
static analyzer been previously examined or used in academic research. There is
also practical relevance. Because the CTWs are related to security---even if
only implicitly, and because a large number of packages is examined, an answer
to RQ.1 gives a coarse but still useful heuristic for deducing about the
potential risks of running a Linux operating system. The answer can be also
portrayed through software engineering; a large amount of CWE-mapped CTWs
implies that a large amount of work is required to improve the situation. From a
perspective of third-party security researchers, on the other hand, CWEs based
on static analysis may help at prioritizing packages for actual vulnerability
discovery. Then, an answer to RQ.2 sheds light on the typical C programming
obstacles faced by Linux developers. If some particularly CWEs are pronounced,
it may be possible to allocate more resources for fixing these issues or
improving instructions and documentation about the issues. Finally, a similar
rationale applies with respect to RQ.3, which is also generally interesting in
that some particular application domains may be particularly prone to
security-related CTWs. Again, an answer to the question allows to also
hypothesize about potential security risks. For instance: if system libraries
are especially prone to CWE-mapped CTWs, the security risks are likely graver
than with less important application domains, such as, say, graphical user
interface~widgets.

The structure of the paper's remainder is simple. As is
common for an empirical paper, materials and methods are first elaborated in
Section~\ref{sec: methods and data}. Results are then presented in the
subsequent Section~\ref{sec: results}. The final Section~\ref{sec: discussion}
presents a conclusion, a discussion about limitations, a few further branches of
related work, and some ideas about further research possibilities.

\section{Methods and Data}\label{sec: methods and data}

\subsection{Methods}

The methodology is based on the static analyzer fairly recently implemented in
the GCC for C (but not C++) code. It is based on coverage-guided symbolic
execution and thus does not operate entirely at the source code level. In
general, coverage-guided symbolic execution, which is used also in some fuzzing
implementations~\cite{Ruohonen19RSDA}, systematically examines potential
execution paths, assuring that even less frequently executed portions are tested
for potential defects.

The static analyzer was initially introduced in the GCC version 10, released in
2020. The rationale behind introducing the static analyzer was based on a
well-grounded argument that bugs should be caught early on, and that a static
analyzer directly embedded to a compiler fits well into the commonplace
edit-compile-debug C development cycle, also reducing the reliance on
additional, possibly commercial third-party tools~\cite{Malcolm20}. This
rationale seems sound already because existing results indicate that integration
of static analyzers into development workflows has sometimes been an
obstacle~\cite{Otetoyan18}. The integration is useful also for research
purposes.

Furthermore, a compiler is a natural place for implementing a static analyzer
already because ``most compilers run many separate static analyses before and
after code is generated''~\cite{Thomson21}. In~line with existing practices,
such as those promoted by the Juliet test suite~\cite{NIST17}, also the CWE
mappings were included already in the initial release. These mappings are likely
important because existing results also indicate that developers have had
difficulties in understanding warnings outputted by static
analyzers~\cite{Baca13}. A~particularly noisy static analyzer with
undecipherable warnings may even increase the probability that developers will
bypass relevant warnings out of annoyance or ignorance~\cite{Baca13,
  Zitser04}. Thus, understandability and help for diagnosis are important design
criteria for static analyzers in general~\cite{Charoenwet24, Nachtigall19}. To
this end, the rationale to include the CWE mappings was that these make ``the
output more clear, improves precision, and gives you something simple to type
into search engines'' \cite{Malcolm20}. In addition to such practical benefits,
the explicit CWE mappings are useful for empirical research purposes, as will be
demonstrated.

The initial version of the static analyzer in the GCC version~10 contained 15
CTWs. Since then, the amount of warnings has grown steadily; the GCC version~13
used in the present paper already contains 47 compile-time warnings in
total~\cite{Malcolm23}. Of these warnings, the paper only considers those
explicitly mapped to CWEs. This choice ensures that software security remains
firmly on the agenda. Because CWEs are about software security weaknesses and
thus often severe, the choice also aligns with previous observations about the
use of static analyzers in software development; many developers prioritize
warnings that are severe~\cite{Vassallo18}. Although some CWEs are difficult to
rigorously define and operationalize in terms of source
code~\cite{GosevaPopstojanova15}, the CWEs considered are all about well-known
bug types. Because potentially obscure bugs are likely therefore excluded, also
the amount of false positives may reduce. Table~\ref{tab: cwes} enumerates all
CWEs considered empirically. For brevity, the listing only includes those CWEs
that were detected via the GCC's CTWs in the popular C packages included in the
dataset. The listing is ordered according to the frequencies of the CWEs in the
dataset.

\begin{table}[th!b]
\centering
\caption{CTW-based CWEs Included in the Dataset}
\label{tab: cwes}
\begin{tabularx}{\linewidth}{lX}
\toprule
CWE & Description \\
\hline
CWE-457 & Use of uninitialized variable \\
CWE-476 & NULL pointer dereference \\
CWE-690 & Unchecked return value to NULL pointer dereference \\
CWE-401 & Missing release of memory after effective lifetime \\
CWE-775 & Missing release of file descriptor or handle after lifetime \\
CWE-131 & Incorrect calculation of buffer size \\
CWE-126 & Buffer over-read \\
CWE-122 & Heap-based buffer overflow \\
CWE-127 & Buffer under-read \\
CWE-121 & Stack-based buffer overflow \\
CWE-686 & Function call with incorrect argument type \\
CWE-685 & Function call with incorrect number of arguments \\
CWE-762 & Mismatched memory management routines \\
CWE-415 & Double free \\
CWE-416 & Use after free \\
CWE-1341 & Multiple releases of same resource or handle \\
CWE-479 & Signal handler use of a non-reentrant function \\
CWE-124 & Buffer underwrite (``buffer underflow'') \\
CWE-590 & Free of memory not on the heap \\
CWE-787 & Out-of-bounds write \\
CWE-674 & Uncontrolled recursion \\
\bottomrule
\end{tabularx}
\end{table}

Finally, in terms of statistical methods, descriptive statistics are used to
answer to the three research questions. In addition, the well-known
non-parametric Kruskal--Wallis test~\cite{Kruskal52} is used to examine
RQ.3. It tests a null hypothesis that the medians of the package categories are
all equal against the alternative that the median of at least one package
category is different. Although the test allows non-equal sample sizes of the
underlying categories, these were scaled by the number of packages containing C
code---a term soon described.

\subsection{Data}\label{sec: data}

The dataset was assembled from all packages that were distributed in the source
code based Gentoo Linux distribution during the start of the data collection in
26 July, 2024. Only the Linux kernel was excluded beforehand. This choice
justifies the popularity term in the paper's title; all packages analyzed can be
assumed to be widely used and hence popular, well-maintained, and supposedly of
relatively high quality.

Together popularity and a programming language are also the most popular choices
for sampling packages~\cite{Tutko22}. Regarding the C language, the choice is
well-justified because most of the packages directly distributed in Linux
distributions have traditionally been written in the language~\cite{Canei16},
and this point likely still holds today. Although the language's flaws are
well-understood and a small movement toward alternatives such as Rust is
happening, C also still retains its appeal in many domains, such as operating
system kernels, low-level system libraries, high-performance computing, and
resource-constrained systems. Furthermore, the choice to cover packages in a
Linux distribution avoids sampling problems that are typical when operating with
more general software hosting services and ecosystems such as
GitHub~\cite{Zeroauli19a}. Obviously, however, not all packages distributed in
Gentoo are written in C, and some packages are written in multiple programming
languages. A simple criterion was used to determine whether C code was involved
in a package: if the GCC was invoked for C code even once, a package was taken
to contain C code. To again improve the consistency of the terminology used, a
term ``packages containing C code'' (PCCCs) is used for such packages. Knowledge
about PCCCs is important because RQ.3 benefits from scaling.

The underlying Gentoo installation and all packages included in the analysis
were kept as vanilla; that is, in particular, so-called USE flags were not used
to tune the packages' features. Because these flags have been a typical reason
for Gentoo-specific compilation failures~\cite{Lienhardt18}, this choice ensures
that some packages were not unnecessarily excluded. In the same vein, only
``stable'' packages were included; that is, masked packages (such as those
marked with a tilde) were excluded. Such masked packages refer to those that
Gentoo developers consider not yet stable enough or otherwise not ready for
production. This choice further reinforces the popularity sampling
assumption. Regarding compiling itself: as optimizations may interfere with some
of the potential warnings~\cite{FSF24a}, all optimizations were turned off for
all packages compiled. In general, optimizations may sometimes eliminate certain
execution paths and variables from an analysis, thus potentially also reducing
the amount of warnings.

Thus, \texttt{-O0~-fanalyzer} were used as the command line options delivered to
the compiler. With these compiling options, the dataset was then assembled
simply by building all packages through looping the Gentoo's package manager
tree over all package categories except \texttt{sys-kernel} and all packages in
the categories.\footnote{~\url{https://packages.gentoo.org/categories}} Given
the large-scale nature of this data collection process, compiling failures and
other errors were not examined. Instead, these were bypassed by letting the
Gentoo's package manager to continue to further packages with the
\texttt{--keep-going} option. That said, only those PCCCs are included in the
dataset that were successfully compiled. Then, the actual quantitative dataset
was put together by parsing the package manager's and GCC's outputs, including
the latter's CTWs in particular. Finally, a brief remark is in order about this
time-consuming large-scale data collection process; it took over a month to
compile through the Gentoo's package tree in a virtual machine running on a
standard personal computer.

\section{Results}\label{sec: results}

\begin{figure}[th!b]
\centering
\includegraphics[width=\linewidth, height=4cm]{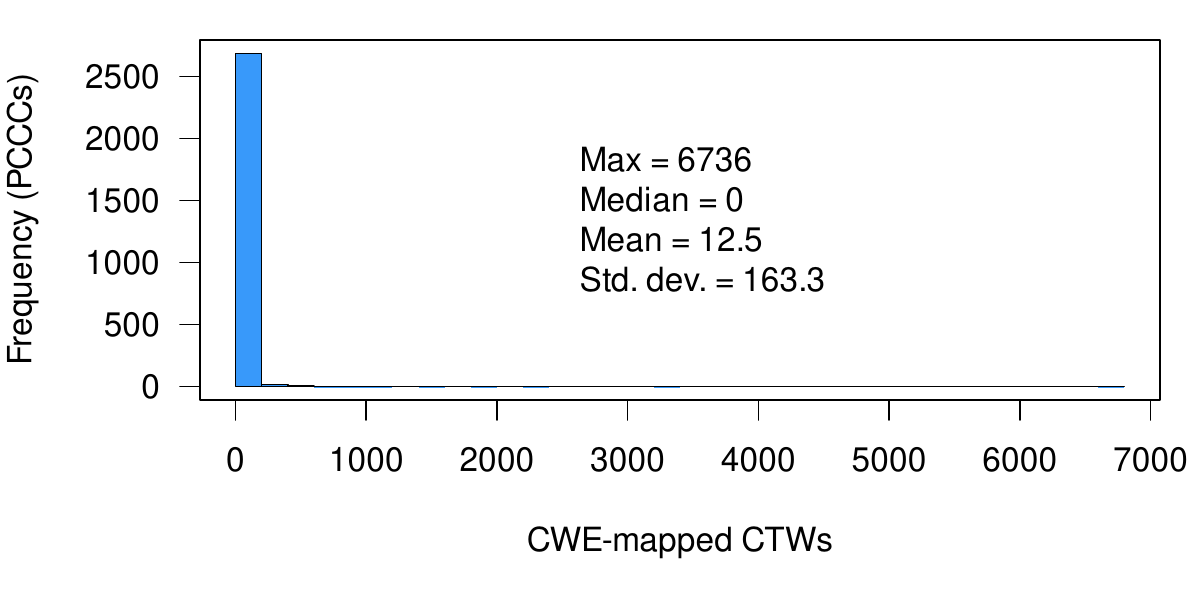}
\caption{A Histogram of the CWEs Across PCCCs}
\label{fig: packages hist}
\end{figure}

The three research questions allow to structure the dissemination of the
empirical results. Thus, RQ.1 asked about the prevalence of CWE-mapped CTWs. For
the $3,538$ successfully compiled PCCCs, a total of $33,817$ compile-time
warnings were detected. At first glance, therefore, a conclusion might be that
more or less security-related CTWs are quite common in popular C packages
distributed in Gentoo. However, such a hasty conclusion is unwarranted because
the CWEs exhibit an extremely long-tailed distribution across the packages, as
can be seen from Fig.~\ref{fig: packages hist}. In other words, a few packages
account for the majority of the CWE-mapped warnings from the GCC's static
analyzer. In fact, the clear majority ($89\%$ to be precise) of the
successfully compiled PCCCs did not exhibit a single CWE-mapped CTW.

Although a long-tailed, power law -like distribution is a classical observation
in the software metrics literature~\cite{Concas07}, the present result does not
necessarily align well with previous studies. For instance, a large-scale static
analysis of Python packages has found that a little below half of Python
packages were prone to warnings from a static analyzer~\cite{Ruohonen21PST}. In
contrast, the present result tells about a few particularly warning-prone
packages. This result supports the motivation for the three RQs presented in the
introduction. While the result does not support a claim that running Linux as
such would be risky, it does align with an argument that installing or deploying
some particular packages may well increase the security risks. Therefore, also
the follow-up motivation is justified; it might be possible to develop CWE-based
risk metrics and other heuristics from the GCC's CTWs. Such metrics and
heuristics might also help third-party security researchers. That is to say, the
results further support a claim that there likely are at least some
``low-hanging fruits'' for vulnerability discovery, even when keeping in mind
the real possibility of false positives.

\begin{figure}[th!b]
\centering
\includegraphics[width=\linewidth, height=10cm]{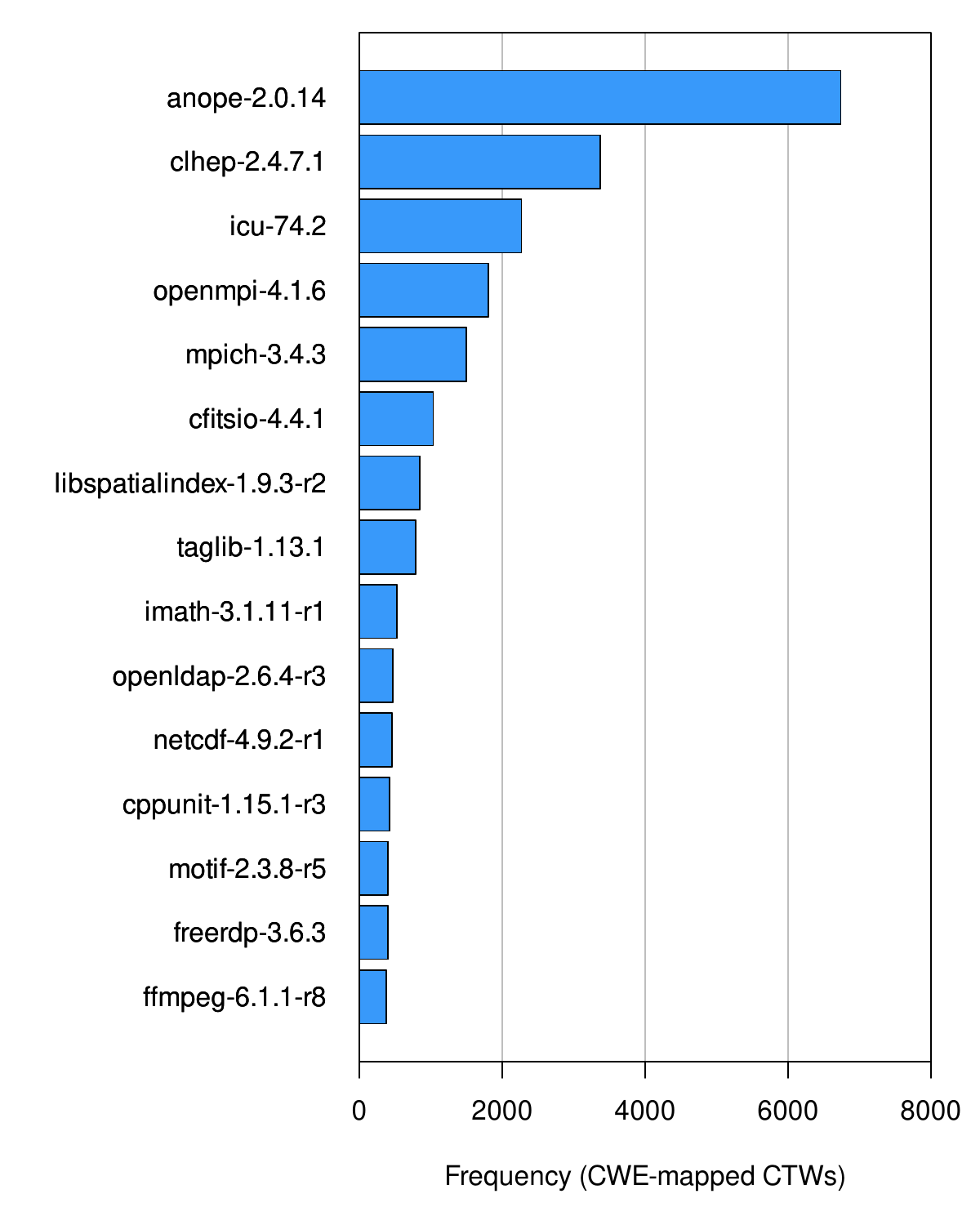}
\caption{Top-15 PCCCs According to CWEs}
\label{fig: packages}
\end{figure}

To take a peek about the actual warning-prone packages, Fig.~\ref{fig: packages}
displays the top-$15$ packages ranked by the CTW-based CWEs. As can be seen,
\texttt{anope}, a suite for different Internet relay chat (IRC) services, leads
the warning scoreboard. In fact, this package alone accounts for as much as
$40\%$ of all CWE-mapped compile-time warnings. While the amount is much
lower, also \texttt{clhep}, a physics library developed by the European
Organization for Nuclear Research (CERN), is quite warning-prone according to
the GCC's static analyzer. The third place is taken by \texttt{icu}, a library
for handling Unicode. Thereafter, the amount of CWEs is much lower, quickly
descending toward warning-free packages. What is important to additionally
remark is that no essential system libraries, core low-level operating system
components, servers, and other generally security-critical software appear in
the \text{top-$15$} CWE-ranking, \texttt{openldap} perhaps notwithstanding.

\begin{figure}[th!b]
\centering
\includegraphics[width=\linewidth, height=10cm]{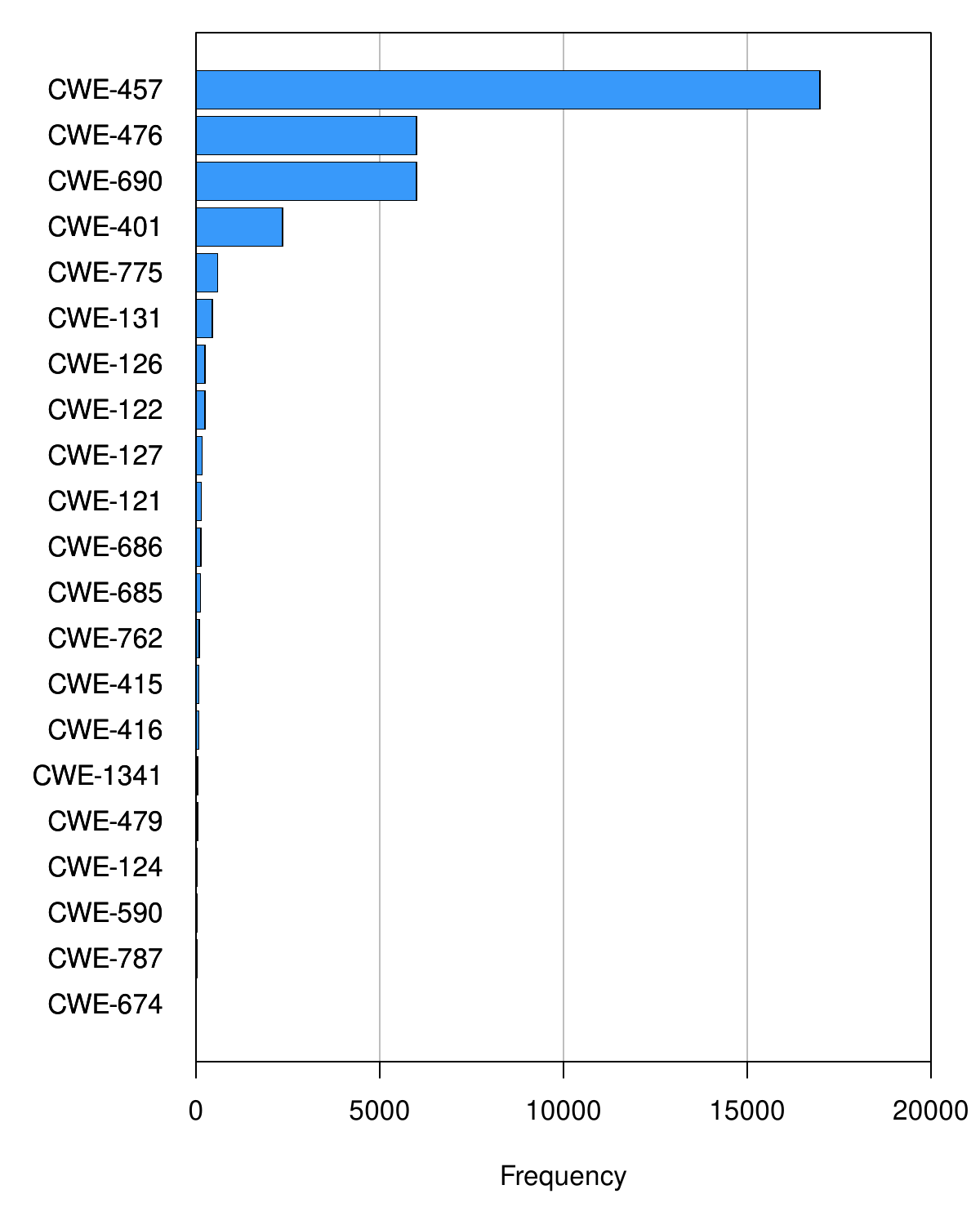}
\caption{CTW-based CWEs Detected}
\label{fig: cwes}
\end{figure}

Turning now to RQ.2, which asked about the prevalence of particular CWEs across
the popular PCCCs. The results are shown in Fig.~\ref{fig: cwes}. When
backtracking to Table~\ref{tab: cwes}, it can be concluded that the use of
uninitialized variables is clearly the most frequent CWE-based programming
mistake group. A~similar result has been obtained also
previously~\cite{Selvaraj23}. While the security concerns with \text{CWE-457}
issues might be often seen as mild, these depend on a context; an uninitialized
variable may become a serious issue if the variable is dereferenced later on. In
general, the security consequences from uninitialized variables may include
information disclosure, bypass of security controls, and even control flow
hijacking~\cite{Milburn17}. Then, the second place is taken by potential NULL
pointer dereference bugs, as captured by CWE-476 and CWE-690. The third place is
occupied by memory and other resource leaks (CWE-401 and CWE-775). Although NULL
pointer dereference issues may manifest themselves as exploitable
vulnerabilities~\cite{Corbet09}, many of the CWEs in Fig.~\ref{fig: cwes} do not
appear or rank low in top rankings of most dangerous software
weaknesses~\cite{Cardoso20}. For instance, classical and serious memory
management issues, such as heap and stack overflows (as captured by CWE-126,
CWE-122, and \text{CWE-121}, among others) are quite rare but still visible as
outliers.

These results beg a question: why are warnings about uninitialized variables,
NULL pointer dereference issues, and resource leaks so common? While NULL
pointer dereference issues can sometimes be difficult to diagnose and debug,
uninitialized variables are arguably rather trivial and mechanical to fix. The
warnings about resource leaks hint that also memory management remains an
issue. Thus, maybe there is something misplaced in existing C programming
language guides and coding styles for the language?  This point is reinforced by
the relatively small amount of security-related memory management warnings,
which oftentimes garner lengthy discussions in the guidelines and coding styles
for the C programming language. Another potential explanation may originate from
the application domains of the warning-prone packages.

The third and final research question asked about potential variance of the
compile-time warnings across application domains. The Gentoo's package
categories are suitable for this task. Before continuing to formal test results,
descriptive statistics are worth looking at beforehand. Thus, Fig.~\ref{fig:
  categories hist} again shows a histogram of the CTW-based CWEs across the
Gentoo's package categories. As can be seen, also this distribution is
long-tailed; $50\%$ of the package categories are without any warnings. However,
the distribution is not as sharp as the earlier one in Fig.~\ref{fig: packages
  hist}, and its tail exhibits also small spikes.

\begin{figure}[th!b]
\centering
\includegraphics[width=\linewidth, height=4cm]{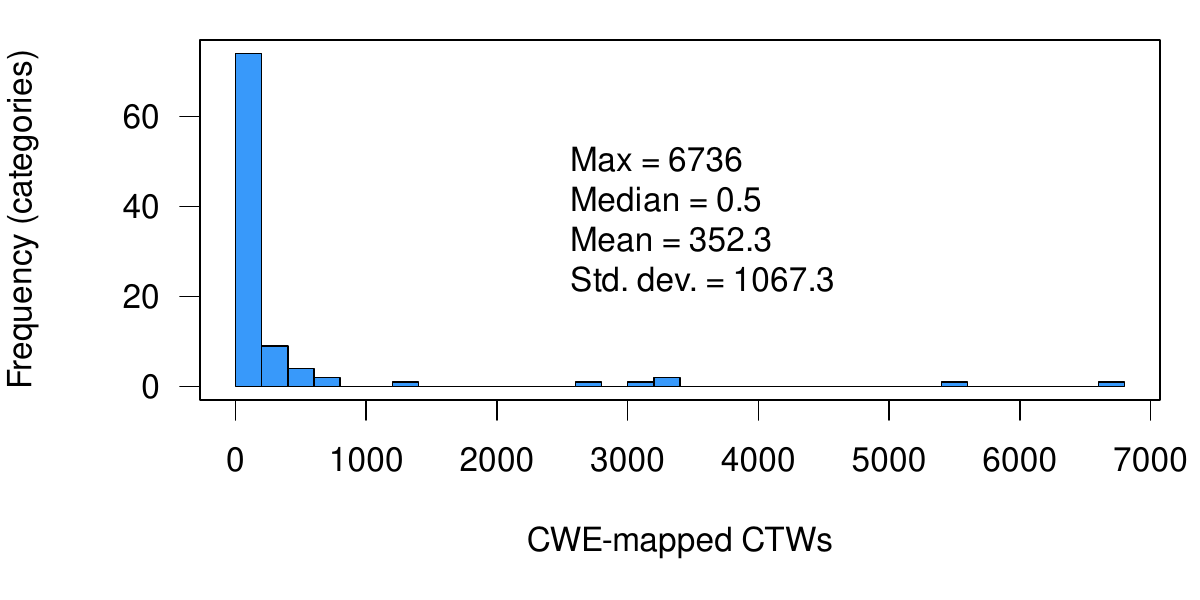}
\caption{A Histogram of the CWEs Across Package Categories}
\label{fig: categories hist}
\end{figure}

\begin{figure}[th!b]
\centering
\includegraphics[width=\linewidth, height=10cm]{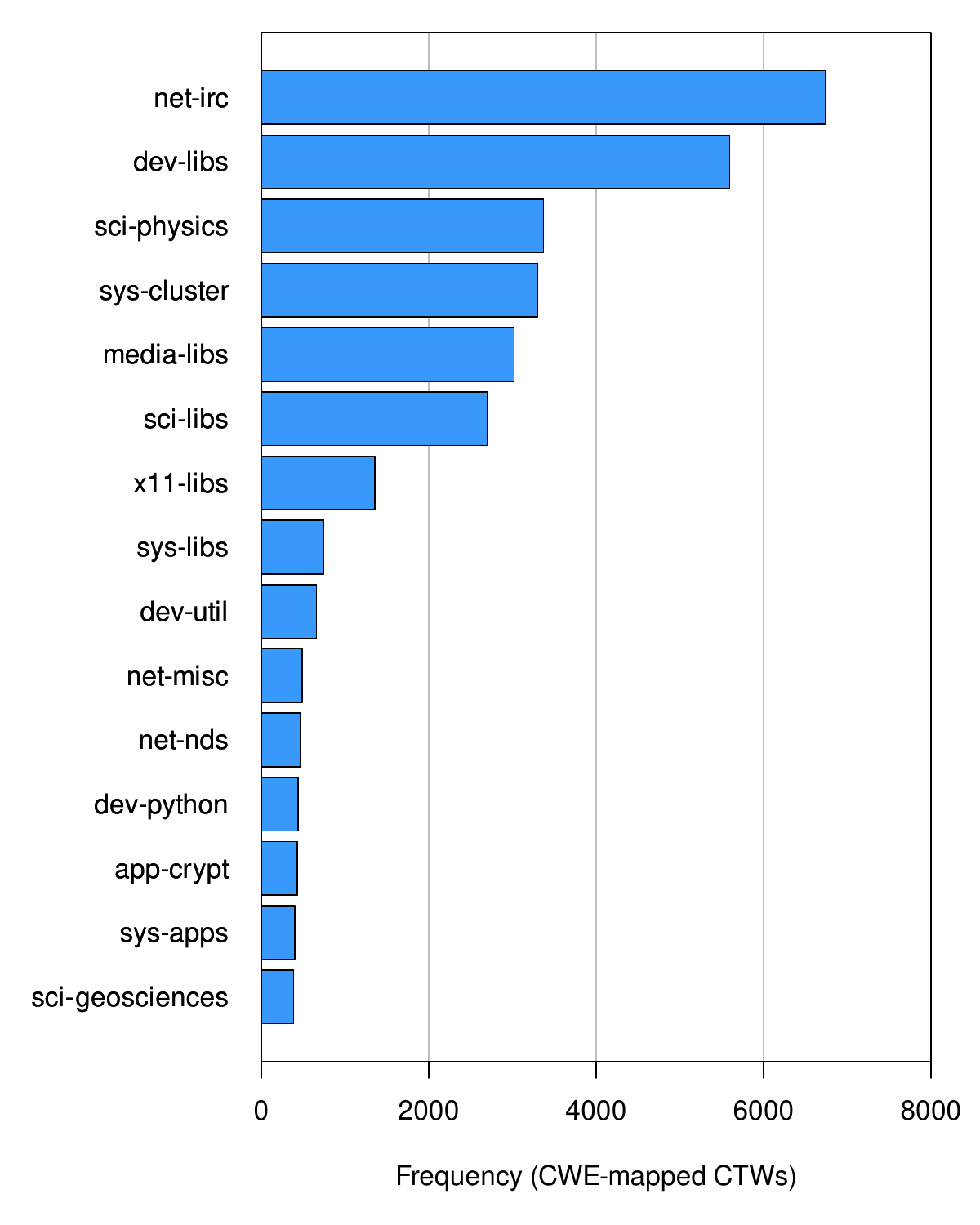}
\caption{Top-15 Package Categories According to CWEs}
\label{fig: categories}
\end{figure}

Then, Fig.~\ref{fig: categories} displays top-$15$ package categories according
to the frequency of CTW-based CWEs. The most frequent category,
\texttt{net-irc}, is entirely because of the noted \texttt{anope} package. The
second place is taken by \texttt{dev-libs}, a category for different software
development libraries. Interestingly, libraries for physics and scientific
libraries in general, as categorized into \texttt{sci-physics} and
\texttt{sci-libs}, also exhibit a relatively large amount of CWE-mapped
compile-time warnings. This observation aligns with a recent study according to
which many machine learning libraries have been quite prone to
vulnerabilities~\cite{Harzevili23}. When compared to the study, however, the
CWEs in Fig.~\ref{fig: cwes} show no numerical issues, \text{CWE-131} perhaps
excluded. While this misalignment may well be because of the GCC analyzer's
capabilities and limitations, the topic seems interesting enough to possibly
examine in further research. Nor do the results in Fig.~\ref{fig: categories}
align well with other large-scale empirical studies. While the Linux kernel was
omitted in the present work, existing studies with Debian's packages indicate
that alongside the kernel, web browsers, email clients, and implementations for
the Java language lead vulnerability rankings~\cite{Alexopoulos18}. Existing
results have also hinted that graphical user interfaces may receive less
development and maintenance effort~\cite{Champion21}. Although the
\texttt{x11-libs} category appears in Fig.~\ref{fig: categories}, none of the
application domains except \texttt{net-irc} and \texttt{dev-libs} pronouncedly
stand out in terms of CTWs. On one hand, this misalignment, in turn, likely
partially reflects differences between warnings from static analyzers and actual
vulnerabilities, including a question about the former's ability to detect the
latter. On the other hand, the question about development and maintenance effort
and its relation to software quality remains open for investigations.

\begin{table}[th!b]
\centering
\caption{Kruskal--Wallis Tests (scaling by PCCCs)}
\label{tab: kw}
\begin{tabular}{lrr}
\toprule
& $\chi^2$ & $p$-value \\
\hline
All packages & $95$ & $0.4807$ \\
\texttt{anope} excluded \qquad\qquad\qquad & $95$ & $0.4807$ \\
\bottomrule
\end{tabular}
\end{table}

The Kruskal-Wallis rank sum test results are shown in Table~\ref{tab: kw}. To
recall: these test a null hypothesis that the medians are all equal across the
Gentoo's package categories. According to the test results, this null hypothesis
remains in force; the medians are equal. Despite the distribution's long tail in
Fig.~\ref{fig: categories hist}, the tail seems to lack power against the half
of the warning-free packages. This results further reinforce the contradiction
with the previous studies mentioned. Having said that, the result is important
in terms of the earlier speculation about risk factors. While keeping in mind
potential false (and true) positives, it seems that users may not particularly
need to worry about some particular application domains being particularly
risky---at least according to a static analyzer.

\begin{figure}[th!b]
\centering
\includegraphics[width=\linewidth, height=8cm]{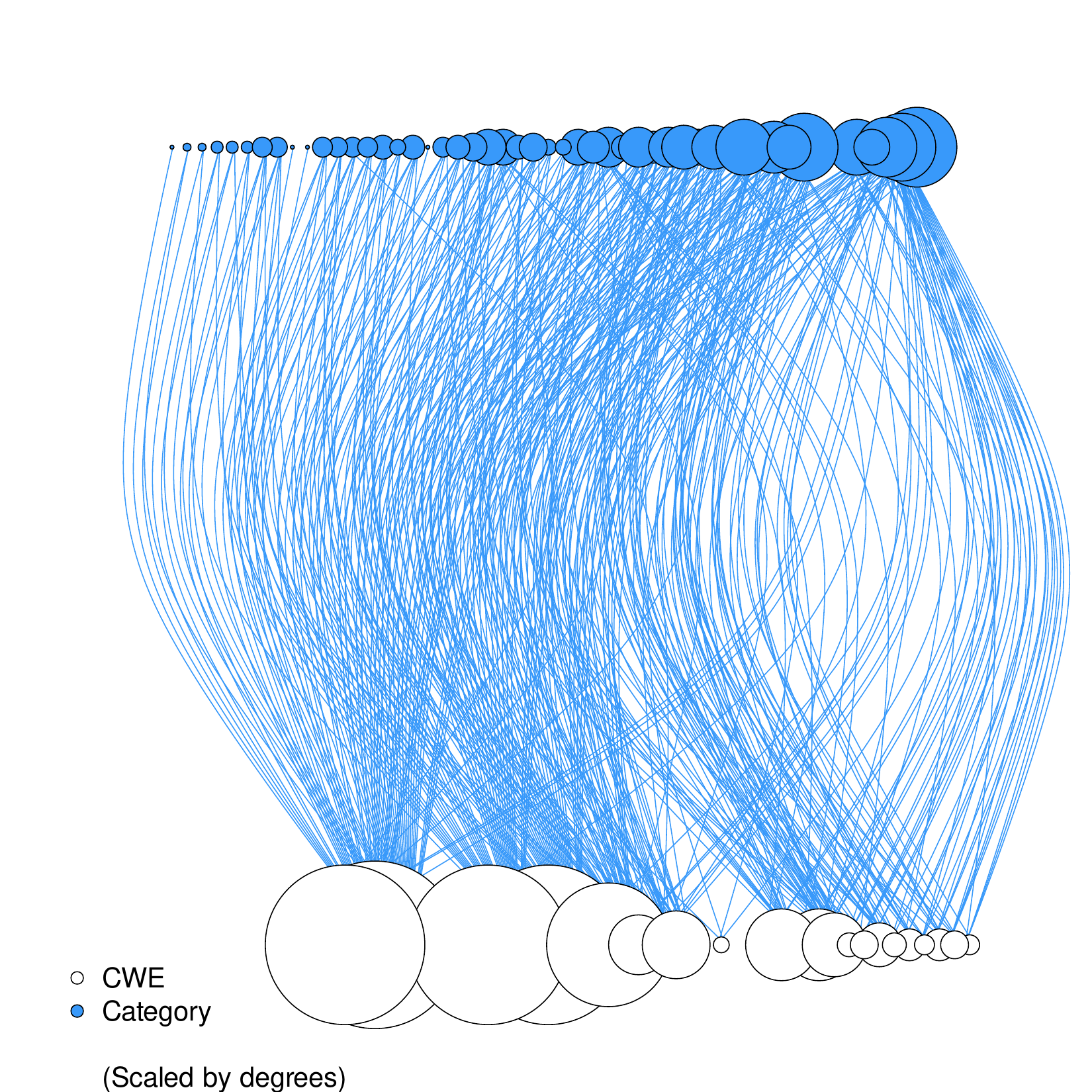}
\caption{Package Categories and CWEs}
\label{fig: categories cwes}
\end{figure}

Although not part of the research questions, it is finally interesting to take a
brief look on whether the distinct CWEs in Fig.~\ref{fig: cwes} also vary across
the package categories. To this end, Fig.~\ref{fig: categories cwes} shows a
bipartite graph between the CWEs and the categories; the former are at the
bottom and the latter at the top, and the sizes of the vertices are scaled by
their degrees. By looking at the CWEs with the highest frequencies---the large
circles colored in white, it can be concluded that these do not match perfectly
the most warning-prone package categories, as defined by the number of CWEs in
the categories. In fact, many of the warning-prone package categories---the
larger circles colored in blue on the top-right---are also associated with the
relatively infrequent CWEs. This observation yields a small caveat to the
reasoning about risks, as it might be that for instance memory management issues
are prevalent in some application domains. While security risks vary across
distinct CWEs, the answer to RQ.3 still holds in terms of the volume of warnings
outputted by the GCC's static analyzer.

\section{Discussion}\label{sec: discussion}

\subsection{Summary}

The paper used a static analyzer recently introduced in the GCC to empirically
examine popular Linux packages fully or partially written in the C programming
language. Three research questions were presented and empirically examined. The
answers to these can be briefly summarized as follows:

\begin{itemize}
\itemsep 5pt
\item{The results indicate that static analysis warnings are common and
  prevalent; over thirty thousand CWE-mapped compile-time warnings were detected
  for about $3,5$ thousand popular C packages. However, the distribution of
  these warnings is highly uneven across the packages examined; the overwhelming
  majority of the packages is entirely without CTWs, while a small minority
  gathers most of the warnings. In other words, outliers dominate.}
\item{The use of uninitialized variables (CWE-457) is clearly the most typical
  source behind the GCC's warnings. NULL pointer dereference issues (CWE-476 and
  CWE-690) take the second place. Rest of the CWEs detected are much less
  common. By implication, software quality might be perhaps improved with
  relatively little amount of effort. It may also be worthwhile to revisit and
  alter coding guidelines and associated practices for the C language, as has
  been suggested also previously \cite{Karapetyants23}. Although actual security
  risks are difficult to evaluate, these might be seen as relatively mild on
  statistical grounds, the apparent outliers perhaps notwithstanding.}
\item{The CWE-mapped compile-time warnings do not statistically vary across
  different application domains. Although there are outliers and likely
  detection errors, no particular application domain seems particularly risky
  for deployments---at least in terms of the static analyzer.}
\end{itemize}

\subsection{Limitations}

\subsubsection{False positives}

foremost, the present work did not consider false positives or other
verification of the GCC analyzer's CTWs. Hence, the security-orientedness of the
results presented remains only implicit and statistical. Nor can anything be
said about true negatives and true positives, that is, whether a package's
source code is bug-free and secure, or the other way around. To patch the issue,
the robustness of the CTW$\mapsto$CWE mappings would first need a closer
examination. For instance, a potential lack of robustness may affect the answer
to RQ.2 in case the GCC is more capable of analyzing some particular CWEs than
others (cf.~\cite{He23}). After a mappings evaluation, actual vulnerabilities
might be subsequently evaluated. It should be also noted that the amount of
false positives likely varies across different GCC versions; hence, further
evaluation work is required also with respect to the compiler's evolution. This
point is also familiar from existing studies~\cite{Reinhold23}. It may also
affect replicability.

\subsubsection{The Linux kernel}

the most important Linux package, the Linux kernel, was omitted from the
analysis. Given the sensitivity and low-level nature of kernel code, this
package would be particularly relevant in terms of false positives and other
potential robustness issues in the GCC's static analyzer.

\subsubsection{Compilation}

regarding internal validity, it may be that some packages force their own
compiler flags, bypassing those specified in Gentoo. While this limitation
affects all three research questions, the statistical effect of such packages is
likely small already due to the large amount of packages covered in the
dataset. Another point is that default configurations defined by Gentoo
developers were used for all packages. While sensible and justifiable due to the
large-scale --- and slow --- data collection process, this choice may cause a
bias because existing studies have shown that static analysis warnings vary
across different configurations~\cite{Medeiros20}. Furthermore, as only
successfully compiled PCCCs were considered, a small bias may be present because
packages that failed to compile may also be more prone to CWE-mapped
compile-time warnings.

\subsubsection{CWEs}

the paper only considered CWE-mapped CTWs, whereas, in reality, also non-mapped
warnings may be relevant for software quality and
security~\cite{Charoenwet24}. While this restriction is justifiable on
statistical and security-orientedness grounds, further work is required to
examine the other warnings, including their potential security consequences.

\subsubsection{Generalizability}

the paper focused on well-known packages distributed in a well-known Linux
distribution. By implication, it is likely that CTWs are more prevalent in less
known and less popular packages. Deploying such packages then likely also
increases the security risks. Eighth, only packages in one Linux distribution
were examined. This limitation is related to the previous point. For instance,
Debian and Ubuntu distribute much more packages than
Gentoo~\cite{Spinellis12}. Hence, in purely statistical terms, these
distributions also likely carry more CTWs, perhaps being less secure \textit{as
  a whole}. Together, these two limitations also pinpoint toward a sharper
analysis of different deployment scenarios involving mixtures of~packages. Here,
CWE-mapped CTWs might work together with existing vulnerability counts and other
software quality metrics to help those who deploy containers or virtual machines
to heuristically deduce about potential security risks in advance. Given the
answer to RQ.3, warnings from a static analyzer might balance existing risk
metrics, giving a more nuanced picture about software quality. In any case, the
topic is generally important because ready-made container images are known to
involve various security issues, including unpatched packages that contain
verified software vulnerabilities~\text{\cite{Sultan19, Zeroauli19b}}. To this
end, security-related CTWs might provide an addition to existing software
security metrics. This point serves to motivate a few ideas about further
research.

\subsection{Related and Further Work}

\subsubsection{Empirical research}

one option for further empirical work would be to enlarge a sample size even
further. As was noted, the limitation to popular packages may cause a bias
because less popular packages might be more prone to security-related CTWs. It
would be relevant to examine this potential bias because existing results
indicate that popular packages actually often have more \textit{reported} bugs
and vulnerabilities, supposedly because more developers and users are reporting
issues about widely used packages than about less popular or even obscure
packages~\cite{Herraiz11, Ruohonen19EASE, Ruohonen25CSA}. To this end, it has
recently been observed and argued that the conventional wisdom from reliability
engineering may not hold well in the open source context; the vulnerability
counts may not decrease as popular packages mature~\cite{Alexopoulos20}. Having
said that, Linux as an operating system lacks a universal hosting service or a
package ecosystem, and thus ecosystem-wide analysis is generally impossible,
unlike with packages specific to programming
languages~\cite{Ruohonen21PST}. Sampling packages from GitHub or other hosting
platforms would also lead to the well-known issues with popularity
metrics~\cite{Zeroauli19a}. Therefore, a more sharply focused comparative
analysis might be an alternative; here, the Berkeley software distributions
(BSDs) provide a classical research setup for empirical
comparisons~\cite{Ruohonen19RSDA, Spinellis09}. Another plausible option would
be a focus on static analyzers rather than~packages, operating systems, or
ecosystems.

\subsubsection{Benchmarks}

benchmarking static analyzers is a classical research topic~\cite{Breuer06,
  Emanuelsson08, NguyenDuc21, Zitser04}. A benchmark would be relevant also for
the developers who maintain and develop the GCC's static analyzer. More
generally, a static analyzer's ability to detect actual vulnerabilities in C
code remains a relevant topic, and existing static analyzers are not performing
particularly well at this task~\text{\cite{Aloraini17, Croft21, Gentsch20,
    GosevaPopstojanova15}}. These results correlate with those from surveys
according to which static analysis is frequently used in (commercial) software
development, but its impact upon improving software security is perceived as
limited~\cite{Rindell21IST}. Furthermore, the results presented have
implications also for benchmark studies because these studies, such as
\cite{Ferrag25}, seem to concentrate on CWEs that are not actually prevalent
according to the results. In fact, many of the top-ranked CWEs in Fig.~\ref{fig:
  cwes} are oftentimes missing.

Given the rationale for introducing the analyzer, a~benchmark might allow to
also deduce about the future prospects in the static analyzer landscape,
including the viability of commercial tools in the future. In terms of the GCC's
direct competitors, further work is also required on the static analysis
capabilities of Clang/LLVM. There is also some existing work in this regard~(see
\cite{Babati17}; and the \cite{Guo23}'s references). Another promising research
path would involve benchmarking of static analyzers against tools for dynamic
analysis. For instance: NULL pointer dereference issues perhaps notwithstanding,
the potential issues detected by the GCC's static analyzer are quite different
from those found via fuzzing~\cite{Ding21, Ruohonen19RSDA}. Therefore, a good
research question would be how these and other software testing techniques
complement each other. It would also be worthwhile to examine whether fixing
warnings from a static analyzer reduces issues found via fuzzing later on.

\subsubsection{Static analysis use}

furthermore, a further classical research topic is the ways open source and
other developers use (and possibly misuse) static analyzers~\text{\cite{Baca13,
    Vassallo20, Vassallo18, Zampetti17}}. Also this line of research would be
relevant for the developers of the GCC's static analyzer. For instance, a good
research question would involve examining how easy it is to comprehend and
analyze the GCC's CTWs. Analogously, it would be important to know whether the
explicit CWE mappings are as useful as presumed. The topic is important because
existing results indicate that many automated software testing techniques,
including those based on symbolic execution, are oftentimes difficult for
developers to apply due to problems in reproducibility, interpretability,
prioritization, cognitive load, and the manual effort
required~\text{\cite{Ghosh25, Molnar24, Otetoyan18, Ruohonen19RSDA,
    Swierzy24}}. If such problems are present also with the GCC's static
analyzer, it might be worthwhile to consider allocating more development
resources and time to improve the analyzer's output rather than extending the
coverage of different bug~types. As recently pointed out~\cite{Charoenwet24},
one potential improvement might be an incorporation of severity information to
the warnings. In fact, since concrete, reported, and archived vulnerabilities,
as typically identified with Common Vulnerabilities and Exposures (CVEs), are
linked to CWEs, it would be possible to use the Common Vulnerability Scoring
System (CVSS) information attached to published CVEs as severity scores. While
alterations are possible, this is essentially also the way the top-CWE rankings
are done~\cite{Cardoso20}. If nothing else, such scores might help developers in
prioritization of warnings.

\subsubsection{Documentation}

alongside integration to other tools and frameworks, better documentation is
what many developers hope for from static analyzers~\cite{Yeboah24}. To this
end, the CWE framework might also allow to improve documentation. In particular,
the framework contains also many so-called views, which are CWE groupings for
different topics and tasks, such as root cause analysis and attack
scenarios. There is also an existing view for secure coding with the C
language~\cite{MITRE24b}. Thus, it might be interesting to evaluate how well
this view or some other view corresponds with the CTW-based CWEs from the GCC's
static analyzer. The work could be extended also to CWEs typically found via
fuzzing and other software testing techniques. Alternatively: to help developers
using static analyzers, it might be reasonable to develop a new view
specifically for CWEs typically found via static analysis.

\subsubsection{Ecosystems}

a further worthwhile but perhaps challenging research avenue would open by
focusing on the larger Linux ecosystem. The present work concentrated on a
so-called downstream distributor, whereas static analysis is presumably more
relevant for the actual upstream developers of the packages considered. However,
there is a link between the two; bugs are typically triaged between upstream and
downstream, although plenty of problems still exist, including with respect to
traceability and associated tools~\cite{Lin22}. The triaging aspects are
important already because patching source code at the distribution-level
(instead of the upstream-level) involves well-known security and other
risks~\cite{Ahmad08, Lin22}. In terms of research questions, it would be
relevant to know whether the downstream Gentoo developers report CTW-based
issues, which may or may not be actual bugs, to upstream developers.

\subsubsection{Dashboards}

as an alternative, therefore, the downstream Gentoo developers might consider a
continuously updated online dashboard for all CWE-mapped CTWs. It has also been
argued and observed that outputs from a static analyzer may predict whether a
software component will soon be difficult and costly to
maintain~\cite{Karanikiotis21}. A dashboard might thus also help Gentoo
developers to determine whether some particular packages might be deprecated
from the distribution due to their real or perceived lack of quality. Although
there may be practical challenges to get developers interested in automatically
generated (CTW) reports~\cite{Ryan23b}, a dashboard might further help at
directing attention and effort to packages that may have problems in
maintenance. The warning-prone scientific libraries are a good example in this
regard. Though, a benchmark would be required beforehand because a dashboard
reporting false positives and noise is not useful to anyone.

\subsubsection{Triaging}

provided that a benchmark would be available for determining the GCC analyzer's
ability to detect vulnerabilities, it would be relevant to examine also the
triaging of vulnerabilities. Such triaging enlarges the scope, as important
vulnerabilities are typically labeled with CVEs. The CVE assignments have also
continued to face problems, including time delays due to coordination problems,
database quality issues, and other problems~\text{\cite{Anwar22, Lin23,
    Ruohonen18IST}}. Furthermore, Gentoo, unlike Debian and some other
distributions, is not a CVE numbering authority; hence, its developers cannot
assign CVEs directly by themselves. This lack of authority implies that the
vulnerability triaging is even more convoluted. That said, progress has been
recently made in the open source software domain via the introduction of a new
database~\cite{Ruohonen25ICTSS}, which has supposedly made triaging, tracking,
and coordination easier. If a dashboard would be designed and implemented, it
could be further contemplated whether the CTWs reported on it could be linked to
the new vulnerability database as references.

\balance
\bibliographystyle{abbrv}
%\bibliography{se}

\begin{thebibliography}{10}

\bibitem{Ahmad08}
D.~Ahmad.
\newblock {T}wo {Y}ears of {B}roken {C}rypto: {D}ebian's {D}ress {R}ehearsal
  for a {G}lobal {PKI} {C}ompromise.
\newblock {\em IEEE Security \& Privacy}, 6(5):70--73, 2008.

\bibitem{Alexopoulos20}
N.~Alexopoulos, S.~M. Habib, , S.~Schulz, and M.~M\"{u}hlh\"{a}user.
\newblock {T}he {T}ip of the {I}ceberg: {O}n the {M}erits of {F}inding
  {S}ecurity {B}ugs.
\newblock {\em ACM Transactions on Privacy and Security}, 24(1):1--33, 2020.

\bibitem{Alexopoulos18}
N.~Alexopoulos, S.~M. Habib, S.~Schulz, and M.~M\"uhlh\"auser.
\newblock {M}-{STAR}: {A} {M}odular, {E}vidence-{B}ased {S}oftware
  {T}rustworthiness {F}ramework.
\newblock {A}rchived manuscript. Available online in July 2024:
  \url{https://arxiv.org/abs/1801.05764}, 2018.

\bibitem{Aloraini17}
B.~Aloraini and M.~Nagappan.
\newblock {E}valuating {S}tate-of-the-{A}rt {F}ree and {O}pen {S}ource {S}tatic
  {A}nalysis {T}ools {A}gainst {B}uffer {E}rrors in {A}ndroid {A}pps.
\newblock In {\em Proceedings of the IEEE International Conference on Software
  Maintenance and Evolution (ICSME 2017)}, pages 295--306, Shanghai, 2017.
  IEEE.

\bibitem{Anwar22}
A.~Anwar, A.~Abusnaina, S.~Chen, F.~Li, and D.~Mohaisen.
\newblock {C}leaning the {NVD}: {C}omprehensive {Q}uality {A}ssessment,
  {I}mprovements, and {A}nalyses.
\newblock {\em IEEE Transactions on Dependable and Secure Computing},
  19(6):4255--4269, 2022.

\bibitem{Ayewah08}
N.~Ayewah, D.~Hovemeyer, J.~D. Morgenthaler, J.~Penix, and W.~Pugh.
\newblock {U}sing {S}tatic {A}nalysis to {F}ind {B}ugs.
\newblock {\em IEEE Software}, 25(5):22--29, 2008.

\bibitem{Babati17}
B.~Babati, G.~Horv\'ath, V.~M\'ajer, and N.~Pataki.
\newblock {S}tatic {A}nalysis {T}oolset with {C}lang.
\newblock In {\em Proceedings of the 10th International Conference on Applied
  Informatics}, pages 23--29, Eger, 2017.
\newblock {A}vailable online in July 2024:
  \url{https://icai.uni-eszterhazy.hu/icai2017/uploads/papers/2017/final/ICAI.10.2017.23.pdf}.

\bibitem{Baca13}
D.~Baca, B.~Carlsson, K.~Petersen, and L.~Lundberg.
\newblock {I}mproving {S}oftware {S}ecurity {W}ith {S}tatic {A}utomated {C}ode
  {A}nalysis in an {I}ndustry {S}etting.
\newblock {\em Software: Practice and Experience}, 43(3), 2013.

\bibitem{Beller16}
M.~Beller, R.~Bholanath, S.~McIntosh, and A.~Zaidman.
\newblock {A}nalyzing the {S}tate of {S}tatic {A}nalysis: {A} {L}arge-{S}cale
  {E}valuation in {O}pen {S}ource {S}oftware.
\newblock In {\em Proceedings of the IEEE 23rd International Conference on
  Software Analysis, Evolution, and Reengineering (SANER 2016)}, pages
  470--481, Osaka, 2016. IEEE.

\bibitem{Breuer06}
P.~T. Breuer and S.~Pickin.
\newblock {O}ne {M}illion ({LOC}) and {C}ounting: {S}tatic {A}nalysis for
  {E}rrors and {V}ulnerabilities in the {L}inux {K}ernel {S}ource {C}ode.
\newblock In {\em Proceedings of the 11th Ada-Europe International Conference
  on Reliable Software Technologies (Ada-Europe 2006)}, pages 56--70, Porto,
  2006. Springer.

\bibitem{Canei16}
M.~Caneill, D.~M. Germ\'an, and S.~Zacchiroli.
\newblock {T}he {D}ebsources {D}ataset: {T}wo {D}ecades of {F}ree and {O}pen
  {S}ource {S}oftware.
\newblock {\em Empirical Software Engineering}, 22:1405--1437, 2016.

\bibitem{Champion21}
K.~Champion and B.~M. Hill.
\newblock {U}nderproduction: {A}n {A}pproach for {M}easuring {R}isk in {O}pen
  {S}ource {S}oftware.
\newblock In {\em Proceedings of the IEEE International Conference on Software
  Analysis, Evolution and Reengineering (SANER 2021)}, pages 388--399,
  Honolulu, 2021. IEEE.

\bibitem{Charoenwet24}
W.~Charoenwet, P.~Thongtanunam, V.~Pham, and C.~Treude.
\newblock {A}n {E}mpirical {S}tudy of {S}tatic {A}nalysis {T}ools for {S}ecure
  {C}ode {R}eview.
\newblock In {\em Proceedings of the 33rd ACM SIGSOFT International Symposium
  on Software Testing and Analysis (ISSTA 2024)}, pages 691--703, Vienna, 2024.
  ACM.

\bibitem{Concas07}
G.~Concas, M.~Marchesi, S.~Pinna, and N.~Serra.
\newblock {P}ower-{L}aws in a {L}arge {O}bject-{O}riented {S}oftware {S}ystem.
\newblock {\em IEEE Transactions on Software Engineering}, 33(10):687--708,
  2007.

\bibitem{Corbet09}
J.~Corbet.
\newblock {F}un {W}ith {NULL} {P}ointers, {P}art 1.
\newblock {L}inux {W}eekly {N}ews {(LWN)}, available online in September 2024:
  \url{https://lwn.net/Articles/342330/}, 2009.

\bibitem{Croft21}
R.~Croft, D.~Newlands, Z.~Chen, and M.~A. Babar.
\newblock {A}n {E}mpirical {S}tudy of {R}ule-{B}ased and {L}earning-{B}ased
  {A}pproaches for {S}tatic {A}pplication {S}ecurity {T}esting.
\newblock In {\em Proceedings of the 15th ACM/IEEE International Symposium on
  Empirical Software Engineering and Measurement (ESEM 2021)}, pages 1--12,
  Bari, 2021. ACM.

\bibitem{Ding21}
Z.~Y. Ding and C.~{Le Goues}.
\newblock {A}n {E}mpirical {S}tudy of {OSS}-{F}uzz {B}ugs.
\newblock In {\em Proceedings of the IEEE/ACM 18th International Conference on
  Mining Software Repositories (MSR 2021)}, pages 131--142, Madrid, 2021. IEEE.

\bibitem{Emanuelsson08}
P.~Emanuelsson and U.~Nilsson.
\newblock {A} {C}omparative {S}tudy of {I}ndustrial {S}tatic {A}nalysis
  {T}ools.
\newblock {\em Electronic Notes in Theoretical Computer Science}, 217:5--21,
  2008.

\bibitem{Fard13}
A.~M. Fard and A.~Mesbah.
\newblock {JSNOSE}: {D}etecting {J}ava{S}cript {C}ode {S}mells.
\newblock In {\em Proceedings of the IEEE 13th International Working Conference
  on Source Code Analysis and Manipulation (SCAM 2013)}, pages 116--125,
  Eindhoven, 2013. IEEE.

\bibitem{Ferrag25}
M.~A. Ferrag, A.~Battah, N.~Tihanyi, R.~Jain, D.~Maimu, F.~Alwahedi,
  T.~Lestable, N.~S. Thandi, A.~Mechri, M.~Debbah, and L.~C. Cordeiro.
\newblock {S}ecure{F}alcon: {A}re {W}e {T}here {Y}et in {A}utomated {S}oftware
  {V}ulnerability {D}etection {W}ith {LLMs}?
\newblock {\em IEEE Transactions on Software Engineering}, 51(4):1248--1265,
  2025.

\bibitem{FSF24a}
{FSF}.
\newblock {O}ptions {T}hat {C}ontrol {S}tatic {A}nalysis.
\newblock {GCC} {O}nline {D}ocumentation, the {F}ree {S}oftware {F}oundation
  {(FSF)}, Inc. Available online in July 2024:
  \url{https://gcc.gnu.org/onlinedocs/gcc/Static-Analyzer-Options.html}, 2024.

\bibitem{Cardoso20}
C.~C. Galhardo, P.~Mell, I.~Bojanova, and A.~Gueye.
\newblock {M}easurements of the {M}ost {S}ignificant {S}oftware {S}ecurity
  {W}eaknesses.
\newblock In {\em Proceedings of the 36th Annual Computer Security Applications
  Conference (ACSAC 2020)}, pages 154--164, Austin, 2020. ACM.

\bibitem{Gentsch20}
C.~Gentsch.
\newblock {E}valuation of {O}pen {S}ource {S}tatic {A}nalysis {S}ecurity
  {T}esting {(SAST)} {T}ools for {C}.
\newblock {G}erman {A}erospace {C}enter, {T}echnical {R}eport
  {DLR}-{IB}-{DW}-{JE}-2020-16. Available online in July 2024:
  \url{https://elib.dlr.de/133945/1/2020_Gentsch_SAST.pdf}, 2020.

\bibitem{Ghosh25}
R.~Ghosh, S.~De, and M.~Mondal.
\newblock ``{I} {W}asn't {S}ure {I}f {T}his {I}s {I}ndeed a {S}ecurity
  {R}isk'': {D}ata-{D}riven {U}nderstanding of {S}ecurity {I}ssue {R}eporting
  in {G}it{H}ub {R}epositories of {O}pen {S}ource npm {P}ackages.
\newblock Archived manuscript, available online in June 2025:
  \url{https://arxiv.org/abs/2506.07728}, 2025.

\bibitem{GosevaPopstojanova15}
K.~Goseva-Popstojanova and A.~Perhinschi.
\newblock {O}n the {C}apability of {S}tatic {C}ode {A}nalysis to {D}etect
  {S}ecurity {V}ulnerabilities.
\newblock {\em Information and Software Technology}, 68:18--33, 2015.

\bibitem{Guo23}
Z.~Guo, T.~Tan, S.~Liu, X.~Liu, W.~Lai, Y.~Yang, Y.~Li, L.~Chen, W.~Dong, and
  Y.~Zhou.
\newblock {M}itigating {F}alse {P}ositive {S}tatic {A}nalysis {W}arnings:
  {P}rogress, {C}hallenges, and {O}pportunities.
\newblock {\em IEEE Transactions on Software Engineering}, 49(12):5154--5188,
  2023.

\bibitem{Harzevili23}
N.~S. Harzevili, J.~Shin, J.~Wang, S.~Wang, and N.~Nagappan.
\newblock {C}haracterizing and {U}nderstanding {S}oftware {S}ecurity
  {V}ulnerabilities in {M}achine {L}earning {L}ibraries.
\newblock In {\em Proceedings of the IEEE/ACM 20th International Conference on
  Mining Software Repositories (MSR 2023)}, pages 27--38, Melbourne, 2023.
  IEEE.

\bibitem{He23}
J.~He, R.~MacQueen, N.~Bombardieri, K.~Ali, J.~R. Wright, and C.~Cifuentes.
\newblock {F}inding an {O}ptimal {S}et of {S}tatic {A}nalyzers {T}o {D}etect
  {S}oftware {V}ulnerabilities.
\newblock In {\em Proceedings of the IEEE International Conference on Software
  Maintenance and Evolution (ICSME 2023)}, pages 463--473, Bogot\'a, 2023.
  IEEE.

\bibitem{Herraiz11}
I.~Herraiz, E.~Shihab, T.~H. Nguyen, and A.~E. Hassan.
\newblock {I}mpact of {I}nstallation {C}ounts on {P}erceived {Q}uality: {A}
  {C}ase {S}tudy on {D}ebian.
\newblock In {\em Proceedings of the Working Conference on Reverse Engineering
  (WCRE 2011)}, pages 219--228, Limerick, 2011. IEEE.

\bibitem{Hjerppe19IFIP}
K.~Hjerppe, J.~Ruohonen, and V.~Lepp\"anen.
\newblock {A}nnotation-{B}ased {S}tatic {A}nalysis for {P}ersonal {D}ata
  {P}rotection.
\newblock In M.~Friedewald, M.~\"Onen, E.~Lievens, S.~Krenn, and S.~Fricker,
  editors, {\em Privacy and Identity Management. Data for Better Living: AI and
  Privacy. Proceedings of the 14th IFIP WG 9.2, 9.6/11.7, 11.6/SIG 9.2.2
  International Summer School}, pages 343--358, Windisch, 2019. Springer.

\bibitem{Karanikiotis21}
T.~Karanikiotis, M.~D. Papamichail, and A.~L. Symeonidis.
\newblock {A}nalyzing {S}tatic {A}nalysis {M}etric {T}rends {T}owards {E}arly
  {I}dentification of {N}on-{M}aintainable {S}oftware {C}omponents.
\newblock {\em Sustainability}, 13(22):12848, 2021.

\bibitem{Karapetyants23}
N.~Karapetyants1 and D.~Efanov.
\newblock {A} {P}ractical {A}pproach to {L}earning {L}inux {V}ulnerabilities.
\newblock {\em Journal of Computer Virology and Hacking Techniques},
  19:409--418, 2023.

\bibitem{Kruskal52}
W.~H. Kruskal and W.~A. Wallis.
\newblock {U}se of {R}anks in {O}ne-{C}riterion {V}ariance {A}nalysis.
\newblock {\em Journal of the American Statistical Association},
  47(260):583--621, 1952.

\bibitem{Lienhardt18}
M.~Lienhardt, F.~Damiani, S.~Donetti, and L.~Paolini.
\newblock {M}ulti {S}oftware {P}roduct {L}ines in the {W}ild.
\newblock In {\em Proceedings of the 12th International Workshop on Variability
  Modelling of Software-Intensive Systems (VAMOS 2018)}, pages 89--96, Madrid,
  2018. ACM.

\bibitem{Lin23}
J.~Lin, B.~Adams, and A.~E. Hassan.
\newblock {O}n the {C}oordination of {V}ulnerability {F}ixes: {A}n {E}mpirical
  {S}tudy of {P}ractices {F}rom 13 {CVE} {N}umbering {A}uthorities.
\newblock {\em Empirical Software Engineering}, 28:1--34, 2023.

\bibitem{Lin22}
J.~Lin, H.~Zhang, B.~Adams, and A.~E. Hassan.
\newblock {U}pstream {B}ug {M}anagement in {L}inux {D}istributions: {A}n
  {E}mpirical {S}tudy of {D}ebian and {F}edora {P}ractices.
\newblock {\em Empirical Software Engineering}, 27:1--41, 2022.

\bibitem{Malcolm20}
D.~Malcolm.
\newblock {S}tatic {A}nalysis in {GCC} 10.
\newblock {R}ed {H}at {D}eveloper {B}log. {A}vailable online in July 2024:
  \url{https://developers.redhat.com/blog/2020/03/26/static-analysis-in-gcc-10},
  2020.

\bibitem{Malcolm23}
D.~Malcolm.
\newblock {I}mprovements to {S}tatic {A}nalysis in the {GCC} 13 {C}ompiler.
\newblock {R}ed {H}at {D}eveloper {B}log. {A}vailable online in July 2024:
  \url{https://developers.redhat.com/articles/2023/05/31/improvements-static-analysis-gcc-13-compiler},
  2023.

\bibitem{Medeiros20}
F.~Medeiros, M.~Ribeiro, R.~Gheyi, L.~Braz, C.~K\"astner, S.~Apel, and
  K.~Santos.
\newblock {A}n {E}mpirical {S}tudy on {C}onfiguration-{R}elated {C}ode
  {W}eaknesses.
\newblock In {\em Proceedings of the XXXIV Brazilian Symposium on Software
  Engineering (SBES 2020)}, pages 193--202, Natal, 2020. ACM.

\bibitem{Milburn17}
A.~Milburn, H.~Bos, and C.~Giuffrida.
\newblock {S}afelnit: {C}omprehensive and {P}ractical {M}itigation of
  {U}ninitialized {R}ead {V}ulnerabilities.
\newblock In {\em Proceedings of the Workshop on Network and Distributed System
  Security (NDSS 2017)}, San Diego, 2017. The Internet Society.

\bibitem{MITRE24a}
{MITRE} et~al.
\newblock {C}ommon {W}eakness {E}numeration: {A} {C}ommunity-{D}eveloped {L}ist
  of {SW} \& {HW} {W}eaknesses {T}hat {C}an {B}ecome {V}ulnerabilities.
\newblock {A}vailable online in July: \url{https://cwe.mitre.org/}, 2024.

\bibitem{MITRE24b}
{MITRE} et~al.
\newblock {CWE} {VIEW}: {W}eaknesses {A}ddressed by the {SEI} {CERT} {C}
  {C}oding {S}tandard.
\newblock {A}vailable online in July:
  \url{https://cwe.mitre.org/data/definitions/1154.html}, 2024.

\bibitem{Molnar24}
A.-J. Molnar, S.~Motogna, D.~Cristea, and D.-F. \c{S}otropa.
\newblock {E}xploring {C}omplexity {I}ssues in {J}unior {D}eveloper {C}ode
  {U}sing {S}tatic {A}nalysis and {FCA}.
\newblock In {\em Proceedings of the 50th Euromicro Conference on Software
  Engineering and Advanced Applications (SEAA 2024)}, pages 407--414, Paris,
  2024. IEEE.

\bibitem{Nachtigall19}
M.~Nachtigall, L.~N.~Q. Do, and E.~Bodden.
\newblock {E}xplaining {S}tatic {A}nalysis -- {A} {P}erspective.
\newblock In {\em Proceedins of the 34th IEEE/ACM International Conference on
  Automated Software Engineering Workshop (ASEW 2019)}, pages 29--32, San
  Diego, 2019. IEEE.

\bibitem{NguyenDuc21}
A.~{Nguyen-Duc}, M.~V. Do, Q.~L. Hong, K.~N. Khac, and A.~N. Quang.
\newblock {O}n the {A}doption of {S}tatic {A}nalysis for {S}oftware {S}ecurity
  {A}ssessment---{A} {C}ase {S}tudy of an {O}pen-{S}ource e-{G}overnment
  {P}roject.
\newblock {\em Computers \& Security}, 111:102470, 2021.

\bibitem{NIST17}
{NIST}.
\newblock {J}uliet {C}/{C}++ 1.3.
\newblock {N}ational {I}nstitute of {S}tandards and {T}echnology ({NIST}).
  Available online in July 2024:
  \url{https://samate.nist.gov/SARD/test-suites/112}, 2017.

\bibitem{OpenSSF24}
{OpenSSF}.
\newblock {C}ompiler {O}ptions {H}ardening {G}uide for {C} and {C}++.
\newblock {O}pen {S}ource {S}ecurity {F}oundation {(OpenSSF)}. {A}vailable
  online in July:
  \url{https://best.openssf.org/Compiler-Hardening-Guides/Compiler-Options-Hardening-Guide-for-C-and-C++.html},
  2024.

\bibitem{Otetoyan18}
T.~D. Oyetoyan, B.~Milosheska, M.~Grini, and D.~S. Cruzes.
\newblock {M}yths and {F}acts {A}bout {S}tatic {A}pplication {S}ecurity
  {T}esting {T}ools: {A}n {A}ction {R}esearch at {T}elenor {D}igital.
\newblock In {\em Proceedings of the 19th International Conference on Agile
  Processes in Software Engineering and Extreme Programming (XP 2018)}, pages
  86--103, Porto, 2018. Springer.

\bibitem{Payne02}
C.~Payne.
\newblock {O}n the {S}ecurity of {O}pen {S}ource {S}oftware.
\newblock {\em Information Systems Journal}, 12:61--78, 2002.

\bibitem{Reinhold23}
A.~M. Reinhold, T.~Weber, C.~Lemak, D.~Reimanis, and C.~Izurieta.
\newblock {N}ew {V}ersion, {N}ew {A}nswer: {I}nvestigating {C}ybersecurity
  {S}tatic-{A}nalysis {T}ool {F}indings.
\newblock In {\em Proceedings of the IEEE International Conference on Cyber
  Security and Resilience (CSR 2023)}, pages 28--35, Venice, 2023. IEEE.

\bibitem{Rindell21IST}
K.~Rindell, J.~Ruohonen, J.~Holvitie, S.~Hyrynsalmi, and V.~Lepp\"anen.
\newblock {S}ecurity in {A}gile {S}oftware {D}evelopment: {A} {P}ractitioner
  {S}urvey.
\newblock {\em Information and Software Technology}, 131:106488, 2021.

\bibitem{Ruohonen19EASE}
J.~Ruohonen.
\newblock {A} {D}emand-{S}ide {V}iewpoint to {S}oftware {V}ulnerabilities in
  {W}ord{P}ress {P}lugins.
\newblock In {\em Proceedings of the 23rd Conference on the Evaluation and
  Assessment in Software Engineering (EASE 2019)}, pages 222--228, Copenhagen,
  2019. ACM.

\bibitem{Ruohonen21PST}
J.~Ruohonen, K.~Hjerppe, and K.~Rindell.
\newblock {A} {L}arge-{S}cale {S}ecurity-{O}riented {S}tatic {A}nalysis of
  {P}ython {P}ackages in {PyPI}.
\newblock In {\em Proceedings of the 18th Annual International Conference on
  Privacy, Security and Trust (PST 2021)}, pages 1--10, Auckland (online),
  2021. IEEE.

\bibitem{Ruohonen25CSA}
J.~Ruohonen and Q.~Ramadan.
\newblock {T}he {P}opularity {H}ypothesis in {S}oftware {S}ecurity: {A}
  {L}arge-{S}cale {R}eplication with {PHP} {P}ackages.
\newblock Archived manuscript, available online in June 2025:
  \url{https://arxiv.org/abs/2502.16670}, 2025.

\bibitem{Ruohonen25ICTSS}
J.~Ruohonen and Q.~Ramadan.
\newblock {T}racing {V}ulnerability {P}ropagation {A}cross {O}pen {S}ource
  {S}oftware {E}cosystems.
\newblock In {\em Proceedings of the 37th International Conference on Testing
  Software and Systems (ICTSS 2025)}, pages 325--332, Limassol, 2026. Springer.

\bibitem{Ruohonen18IST}
J.~Ruohonen, S.~Rauti, S.~Hyrynsalmi, and V.~Lepp\"anen.
\newblock {A} {C}ase {S}tudy on {S}oftware {V}ulnerability {C}oordination.
\newblock {\em Information and Software Technology}, 103:239--257, 2018.

\bibitem{Ruohonen19RSDA}
J.~Ruohonen and K.~Rindell.
\newblock {E}mpirical {N}otes on the {I}nteraction {B}etween {C}ontinuous
  {K}ernel {F}uzzing and {D}evelopment.
\newblock In {\em {P}roceedings of the {IEEE} {I}nternational {S}ymposium on
  {S}oftware {R}eliability {E}ngineering {W}orkshops ({ISSREW} 2019)}, pages
  276--281, Berlin, 2019. IEEE.

\bibitem{Ryan23a}
I.~Ryan, U.~Roedig, and K.~Stol.
\newblock {M}easuring {S}ecure {C}oding {P}ractice and {C}ulture: {A} {F}inger
  {P}ointing at the {M}oon is not the {M}oon.
\newblock In {\em Proceedings of the IEEE/ACM 45th International Conference on
  Software Engineering (ICSE 2023)}, pages 1622--1634, Melbourne. IEEE.

\bibitem{Ryan23b}
I.~Ryan, U.~Roedig, and K.~Stol.
\newblock {U}nhelpful {A}ssumptions in {S}oftware {S}ecurity {R}esearch.
\newblock In {\em Proceedings of the 2023 ACM SIGSAC Conference on Computer and
  Communications Security (CCS 2023)}, pages 3460--3474, Copenhagen, 2023. ACM.

\bibitem{Selvaraj23}
M.~Selvaraj and G.~Uddin.
\newblock {A} {L}arge-{S}cale {S}tudy of {IoT} {S}ecurity {W}eaknesses and
  {V}ulnerabilities in the {W}ild.
\newblock {A}rchived manuscript, available online in September 2024:
  \url{https://arxiv.org/abs/2308.13141}, 2023.

\bibitem{Spinellis12}
D.~Spinellis.
\newblock {P}ackage {M}anagement {S}ystems.
\newblock {\em IEEE Software}, 29(2):84--86, 2012.

\bibitem{Spinellis09}
D.~Spinellis, G.~Gousios, V.~Karakoidas, P.~Louridas, P.~J. Adams,
  I.~Samoladas, and I.~Stamelos.
\newblock {E}valuating the {Q}uality of {O}pen {S}ource {S}oftware.
\newblock {\em Electronic Notes in Theoretical Computer Science}, 233:5--28,
  2009.

\bibitem{Sultan19}
S.~Sultan, I.~Ahmad, and T.~Dimitriou.
\newblock {C}ontainer {S}ecurity: {I}ssues, {C}hallenges, and the {R}oad
  {A}head.
\newblock {\em IEEE Access}, 7:52976--52996, 2019.

\bibitem{Swierzy24}
B.~Swierzy, F.~Boes, T.~Pohl, C.~Bungartz, and M.~Meier.
\newblock {S}o{K}: {A}utomated {S}oftware {T}esting for {TLS} {L}ibraries.
\newblock In {\em Proceedings of the 19th International Conference on
  Availability, Reliability and Security (ARES 2024)}, pages 1--12, Vienna,
  2024. ACM.

\bibitem{Thomson21}
P.~Thomson.
\newblock {S}tatic {A}nalysis: {A}n {I}ntroduction.
\newblock {\em ACM Queue}, 19(4):1--13, 2021.

\bibitem{Tutko22}
A.~Tutko, A.~Z. Henley, and A.~Mockus.
\newblock {H}ow {A}re {S}oftware {R}epositories {M}ined? {A} {S}ystematic
  {L}iterature {R}eview of {W}orkflows, {M}ethodologies, {R}eproducibility, and
  {T}ools.
\newblock {A}rchived manuscript. Available online in July 2024:
  \url{https://arxiv.org/abs/2204.08108}, 2022.

\bibitem{Vassallo20}
C.~Vassallo, S.~Panichella, F.~Palomba, S.~Proksch, H.~C. Gall, and A.~Zaidman.
\newblock {H}ow {D}evelopers {E}ngage with {S}tatic {A}nalysis {T}ools in
  {D}ifferent {C}ontexts.
\newblock {\em Empirical Software Engineering}, 25(2):1419--1457, 2020.

\bibitem{Vassallo18}
C.~Vassallo, S.~Panichella, F.~Palomba, S.~Proksch, A.~Zaidman, and H.~Gall.
\newblock {C}ontext {I}s {K}ing: {T}he {D}eveloper {P}erspective on the {U}sage
  of {S}tatic {A}nalysis {T}ools.
\newblock In {\em Proceedings of the IEEE 25th International Conference on
  Software Analysis, Evolution and Reengineering (SANER 2018)}, pages 38--49,
  Campobasso, 2008. IEEE.

\bibitem{West08}
J.~West.
\newblock {S}ecure {P}rogramming with {S}tatic {A}nalysis.
\newblock In {\em {OWASP}-{D}ay {II}}, Rome, 2008. The OWASP Foundation.
\newblock {T}he {O}pen {W}orldwide {A}pplication {S}ecurity {P}roject
  {(OWASP)}. {A}vailable online in July 2024:
  \url{https://wiki.owasp.org/images/a/a9/Owaspday2West.pdf}.

\bibitem{Yeboah24}
J.~Yeboah and S.~Popoola.
\newblock {U}ncovering {U}ser {C}oncerns and {P}references in {S}tatic
  {A}nalysis {T}ools: {A} {T}opic {M}odeling {A}pproach.
\newblock In {\em Proceedings of the 2nd International Conference on Artificial
  Intelligence, Blockchain, and Internet of Things (AIBThings 2024)}, pages
  1--6, Mt Pleasant, 2024. IEEE.

\bibitem{Zampetti17}
F.~Zampetti, S.~Scalabrino, R.~Oliveto, G.~Canfora, and M.~{Di Penta}.
\newblock {H}ow {O}pen {S}ource {P}rojects {U}se {S}tatic {C}ode {A}nalysis
  {T}ools in {C}ontinuous {I}ntegration {P}ipelines.
\newblock In {\em Proceedings of the IEEE/ACM 14th International Conference on
  Mining Software Repositories (MSR 2017)}, pages 334--344, Buenos Aires, 2017.
  IEEE.

\bibitem{Zeroauli19b}
A.~Zerouali, T.~Mens, G.~Robles, and J.~{Gonzalez-Barahona}.
\newblock {O}n the {R}elation {B}etween {O}utdated {D}ocker {C}ontainers,
  {S}everity {V}ulnerabilities and {B}ugs.
\newblock In {\em Proceedings of the 26th International Conference on Software
  Analysis, Evolution and Reengineering (SANER 2019)}, pages 491--501,
  Hangzhou, 2019. IEEE.

\bibitem{Zeroauli19a}
A.~Zerouali, T.~Mens, G.~Robles, and J.~M. {Gonzalez-Barahona}.
\newblock {O}n the {D}iversity of {S}oftware {P}ackage {P}opularity {M}etrics:
  {A}n {E}mpirical {S}tudy of {npm}.
\newblock In {\em Proceedings of the IEEE 26th International Conference on
  Software Analysis, Evolution and Reengineering (SANER 2019)}, pages 589--593,
  Hangzhou, 2019. IEEE.

\bibitem{Zitser04}
M.~Zitser, R.~Lippmann, and T.~Leek.
\newblock {T}esting {S}tatic {A}nalysis {T}ools {U}sing {E}xploitable {B}uffer
  {O}verflows from {O}pen {S}ource {C}ode.
\newblock In {\em Proceedings of the 12th ACM SIGSOFT Twelfth International
  Symposium on Foundations of Software Engineering (FSE 2004)}, pages 97--106,
  Newport Beach, 2004. ACM.

\end{thebibliography}

\end{document}